\documentclass[iop,apj]{emulateapj}

\usepackage{graphicx}	
\usepackage{amsmath}	
\usepackage{amssymb}	
\usepackage{units}
\usepackage{enumitem}
\usepackage{bm}
\usepackage{color}
\usepackage[breaklinks,colorlinks,citecolor=blue,linkcolor=red]{hyperref} 

\def\msun{{\rm\,M_\odot}}

\def\msun{{\rm\,M_\odot}}

\newcommand{\kms}{\, {\rm km\, s}^{-1}}

\newcommand{\be}{\begin{equation}}
\newcommand{\ee}{\end{equation}}

\def\h2{${\rm\,H_2}$}

\newcommand{\beq}{\begin{equation}}
\newcommand{\beqa}{\begin{eqnarray}}
		 \newcommand{\eeq}{\end{equation}}
\newcommand{\eeqa}{\end{eqnarray}}

\begin{document}

\title{Formation of GW190521 via gas accretion onto Population III stellar black hole remnants born in high-redshift minihalos}
\author{Mohammadtaher Safarzadeh\altaffilmark{1,2} \& Zolt\'an Haiman\altaffilmark{3}}
\affil{$^1$Department of Astronomy \& Astrophysics, University of California, Santa Cruz, CA 95064, USA; \href{mailto:msafarza@ucsc.edu}{msafarza@ucsc.edu}}
\affil{$^2$Center for Astrophysics | Harvard \& Smithsonian, 60 Garden Street, Cambridge, MA}
\affil{$^3$Department of Astronomy, Columbia University, New York, NY 10027, USA; \href{mailto:zoltan@astro.columbia.edu}{zoltan@astro.columbia.edu}}

\begin{abstract}
The recent gravitational wave merger event, GW190521, has challenged our understanding of stellar-mass black hole (BH) formation. 
The primary and secondary BH are both inferred to fall inside the pair-instability (PI) mass gap.
Here we propose that the formation of such binaries is possible through gas accretion onto the
BH remnants of Population III (Pop~III) stars born in high-redshift ($z>10$) minihalos.  Once the parent halo has grown to the atomic-cooling limit, even brief episodes of gas accretion in the dense central regions of the halo can
increase the masses of Pop~III remnant BHs above the PI limit. Starting with a BBH with an initial mass of O(100) ${\rm M_{\odot}}$ we find that it would only need to spend about 100~Myr in the inner few pc of an atomic-cooling halo to accrete about 50~${\rm M_{\odot}}$ of material and resemble a system similar to GW190521. The dynamical friction timescale for the binary to sink to the dense inner region of its parent halo is comparable or shorter than the accretion timescale required to increase their mass above the PI limit. Once in the core of the halo, the binary can enter a phase of hyper-Eddington accretion, where it would only take a few thousand years to exceed the PI limit through accretion.  Even more massive BBHs could form through this channel, and be detectable by detectors with improved low-frequency sensitivity.  Single Pop~III BH remnants would also grow through  accretion and could later form binaries dynamically.
As little as a few percent of Pop~III BH remnants may be sufficient to match the rate of massive BBH mergers inferred from GW190521 of $0.13^{+0.3}_{-0.11}\rm Gpc^{-3} yr^{-1}$. 
\end{abstract}


\section{Introduction}
\label{sec:intro}
The recent discovery of the binary black hole (BBH) merger event, GW190521, by the LIGO/Virgo Collaboration (LVC) has challenged theoretical expectations of the formation of BHs via stellar collapse \citep{Abbott:2020dz,Abbott:2020gz}.
With the primary and secondary mass of $85^{+21}_{-14}\msun$, and $66^{+17}_{-18}\msun$, both components of this BBH have masses above the pair-instability (PI) limit \citep{Woosley:2017dj}. 
Although mergers of first-generation BHs can lead to the formation of such massive BHs, the subsequent "hierarchical" merger of two of these BHs would require extremely dense environments.
These environments are rare but could be realized in denser stellar clusters~\citep{Perna2019,Fragione:2020uh,Rodriguez2020,Rizzutto+2020} or in accretion disks~\citep{McKernan+2012,Tagawa2016,Tagawa:2019tu,Yang+2019} in galactic nuclei. Other theories for the formation of such masive BBHs range from changes in nuclear reaction rates \citep[e.g.,][]{Farmer2020}, accretion of gas in protoglobular clusters \citep{Roupas2019}, gas accretion onto primordial black holes \citep{Luca2020} to physics beyond the Standard Model \citep{Sakstein2020}.

Population III (Pop III) stars are stars with zero metallicity formed from pristine gas at the highest redshifts ($z\gtrsim 10$). Due to the lack of metals, fragmentation is suppressed in the primordial gas, and massive stars are expected to be born \citep[e.g.][]{Nakamura2001,Clark2011}.
These stars are expected to have an initial mass function (IMF) that is much flatter \citep[e.g.][]{Hirano+2015}  than the IMF in the low-redshift universe \citep{Salpeter1955}.  They are also expected to be born frequently as binaries \citep[e.g., ][]{Sana2012,Turk+2009,Hirano2018,Chon+2019}. The compact-object remnants of such stars would be detectable with current ground-based gravitational-wave detectors~\citep{Kinugawa+2014,Inayoshi:2016du}.

Pop~III stellar binaries have recently been proposed as possible sources for remnant masses exceeding the PI limit, due to their modified stellar evolution~\citep{Farrell+2020, Kinugawa+2020}. In this \emph{Letter}, we propose a possible alternative solution to the puzzlingly large masses of the BHs seen in GW190521: the collapse of binary Pop~III stars leave behind binary BHs (BBHs) which can accrete gas from the dense inner parts of their parent halos and increase their masses above the PI limit. 

The rapid growth of (single) Pop~III remnant BHs\footnote{In reality, this scenario also applies to remnants of Pop~II stars, too, as long as they have a flat IMF and leave BH remnants not far below the PI limit.} via accretion have long been discussed as a possible pathway for the formation of supermassive ($M\gtrsim 10^9~{M_\odot}$) BHs or their intermediate-mass seeds~\citep[see, e.g., the recent review by][and references therein]{Inayoshi+2020}.
The growth to such large masses is generally disfavored due to the effects of low ambient density caused by radiative feedback~\citep{JohnsonBromm2007,Alvarez+2009} and by supernova explosions of the progenitors in the minihalo~\citep{Whalen+2008}, and also because the BHs generally orbit away from the dense regions \citep{Smith+2018,Pfister+2019}.  In the present context of GW190521, only a much more modest growth -- a mere $\sim$doubling of the masses of the original remnant BHs -- is required.  Here we argue that such modest growth could be naturally achieved via accretion.

\section{Bondi-Hoyle Accretion on to a BBH}

A BH with mass $M_{\rm bh}$ placed inside a gaseous medium with density $\rho_g$ will accrete at the Bondi–Hoyle–Lyttleton \citep[BHL, ][]{Bondi:1952fc,Bondi:1944gc,Hoyle:1939fl} rate given by:
\be
\dot{M}_{\rm bh}=\frac{4\pi G^2 M_{\rm bh}^2 \rho_{g}}{(c_s^2+v_{\rm rel}^2)^{3/2}},
\ee
where $c_s$ is the sound speed in the medium and $v_{\rm rel}$ is the relative speed of the BH and the surrounding medium.  Throughout this work, we treat the BBH as a single point mass, which is justified in the limit that the binary separation is much smaller than the Bondi radius~\citep[e.g., ][]{Farris+2010,Antoni2019}, as in our case (see below).

We assume a $\lesssim$ 100 $\msun $ BBH is born as the remnant of a binary Pop~III star in a minihalo at high redshift  $z\gtrsim 15$.   The Pop~III stellar
progenitor (or other Pop~III star(s) in the same minihalo)
can irradiate and evaporate the ionized gas from the minihalo~\citep{Kitayama+2004,Whalen+2004}, and the remnant BH is likely to find itself in a very low-density environment, unable to accrete efficiently.  However, efficient accretion can commence once the host halo builds up to larger masses, after a delay of order $\sim100$Myr~\citep[e.g.][]{JohnsonBromm2007}. Specifically, here we assume that this happens once the BH's host halo reaches the atomic-cooling threshold, i.e., a halo mass between $10^{7-8}\msun$ at $z\sim 10$. The gas in these halos cools efficiently via atomic H, builds up a concentrated density profile, and the escape velocity from such halos exceeds the $\sim 10~{\rm km~s^{-1}}$ corresponding to the temperature of photoionized ionized gas. Following \citet{Ryu:2016ih}, we adopt a gas density profile inside such a halo parametrized by:
\be
n_g(r)=\frac{n_c}{1+(r/r_c)^2},
\ee
where $n_c$ and $r_c$ are the core particle number density and radius of the halo. We adopt values of $n_c=2.5\times10^{10}\rm cm^{-3}$ and $r_c=0.003$ pc, which is consistent with high-resolution numerical simulations of atomic cooling halos at high redshifts \citep[e.g.][]{Shang+2010,Regan:2014fj,Wise+2019}. We conservatively assume a constant temperature  $T_g=10^4~K$  for the gas (in reality, gas in the inner regions can be cooler once molecular or metal cooling is activated and would yield higher accretion rates). We further adopt $v_{\rm rel}=10\kms$, corresponding to a typical orbital velocity in an atomic cooling halo. 

\section{Accretion timescale to make a GW190521-like BBH}

\begin{figure}
\hspace{-0.2in}
\centering
\includegraphics[width=\columnwidth]{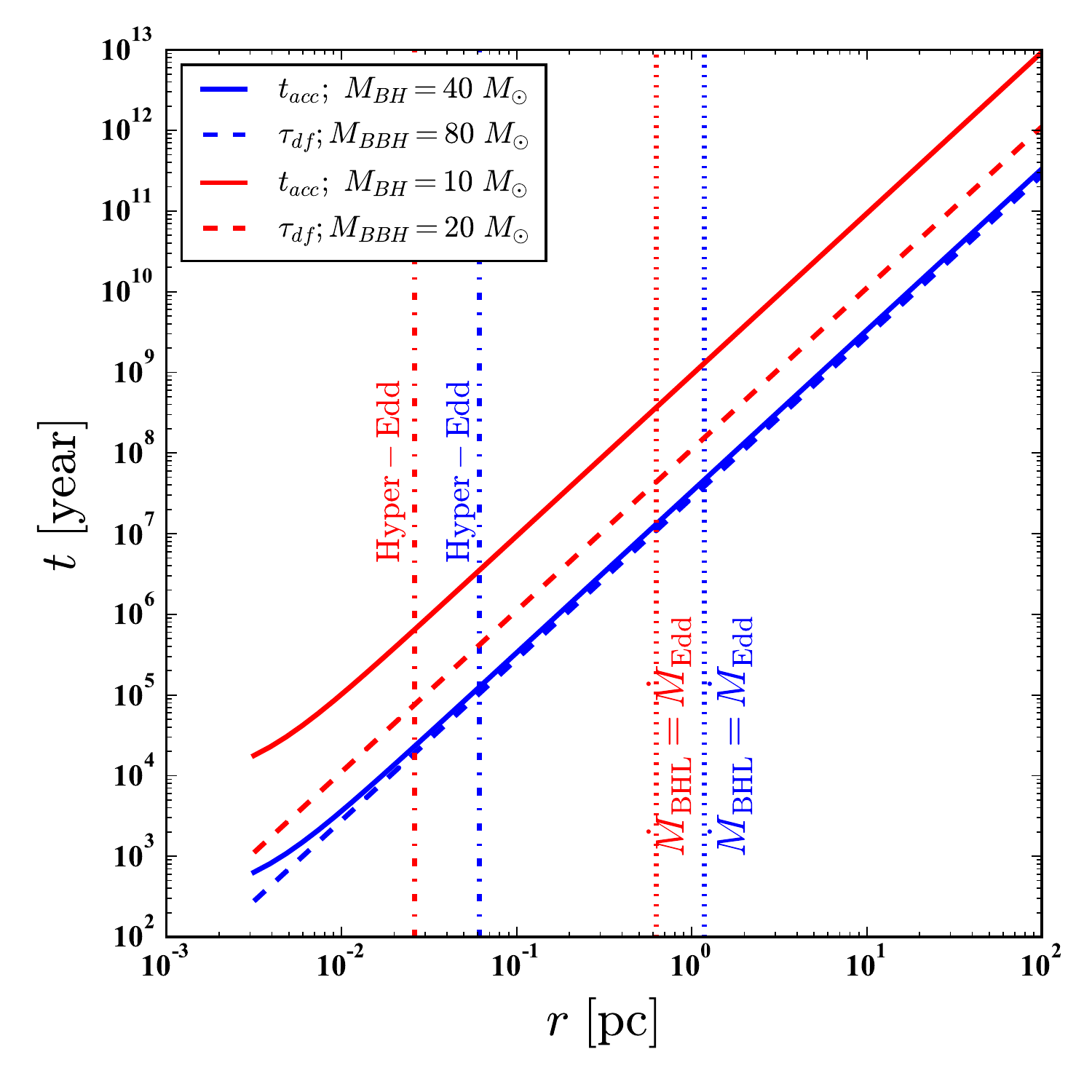}
\caption{The timescale over which a BBH can accrete enough material to resemble a system similar to GW190521. We assume the BBH is traveling at $v_{\rm rel}=10~\kms$ inside a halo with gas temperature of $10^4~{\rm K}$.
This timescale depends on the density in which the BBH resides.
We show two cases for BBHs with initial total masses of 80 and 20 $\msun$. The 80 $\msun$ BBH requires to spend about 100 Myr in the inner 3 pc of the atomic cooling halo to increase its total mass to 150 $\msun$ while the BBH with an initial total mass of 20 $\msun$ would need to spend several Gyrs at this radius to reach 150 $\msun$.
The dashed lines show the dynamical friction timescale for a BBH with a total mass of 20 and 80 $\msun$.  For the lower-mass BBH, this timescale is shorter than the accretion timescale required to bring the component masses of the BBHs to resemble that of GW190521, while for the more massive BBH the timescales become comparable.
The
dotted vertical lines indicate the radius at which the BHL accretion rate exceeds the Eddington limit for each of the BBHs. The dot-dashed lines indicate the radius inside which the BHL accretion enters exceeds $500~\dot{M}_{\rm Edd}$ and allows a phase of hyper-Eddington accretion at this rate~\citep{Inayoshi+2016}.
}
\label{fig_1}
\end{figure}

The timescale over which a BBH increases its total mass to resemble GW190521 (taken to be $150~{\rm M_\odot}$) depends on its birth mass as well as on the density and temperature of the ambient medium. Figure~\ref{fig_1} shows this timescale as a function of the galactocentric radius of an atomic cooling halo for two cases of BBHs with equal component masses of 10 and 40 $\msun$ each (therefore, the total mass of the BBH is 20 and 80 $\msun$, respectively).

As this figure shows, a BBH composed of two BHs with mass 40 $\msun$ that are below the PI limit would need to spend only about 100 Myr in the inner O(1) pc of an atomic cooling halo to increase its component masses each by 35 $\msun$ to make the final total BBH mass about 150 $\msun$. Note that this fiducial distance of a $\sim$pc is comfortably far from the core of the halo. The density at this radius is $\sim 10^5~{\rm cm^{-3}}$, enclosing $\sim 10^5~{\rm M_\odot}$ of gas, i.e., a few percent of the total gas mass in the halo.
A smaller BBH with a total mass of 20 $\msun$ would, on the other hand, need to spend about 10 Gyr within a sub-parsec scale from the center of such a halo to accrete enough mass to increase its component masses to become a BBH with a total mass of 150 $\msun$. 

Would a massive BBH actually sink into the dense regions of the halo? 
A BH with mass $M_{\rm bh}$ would experience a drag force due to both the dark matter and gas in the halo in the direction of opposite its velocity vector, which will result in the sinking of the BH in the center of the halo. In our case, the BHL radius for the BBH's mass is $2GM_{\rm bh}/(c_s^2+v_{\rm rel}^2)\sim 10^{-2}$pc. If the binary separation is much smaller than this value, its center-of-mass should
experience a drag similar to a single point mass~\citep{Antoni2019}.
The drag force due to the gas dominates over that of dark matter \citep{Ryu:2016ih} and is given by:
\be
\vec{a}_{\rm df}=-4\pi G M_{\rm bh} \rho_g(r) \frac{1}{v_{\rm bh}^3}\times f(\mathcal{M})~\vec{v}_{\rm bh},
\ee
with the parametrization of the gas force on the Mach number ($\mathcal{M}$) given in \citet{Ryu:2016ih}. The corresponding dynamical friction (DF) timescale can be computed as $\tau_{\rm df}\propto v/a_{\rm df}$ which results in
\be
\tau_{\rm df}(r)\approx 2 \times 10^4 \left(\frac{M_{\rm bh}}{M_{\odot}}\right)^{-1} \left[1+ \left(\frac{r}{r_c}\right)^2\right] \rm \, yr,
\ee
with the prefactor assuming $\mathcal{M}=1$.
The dashed lines in Figure \ref{fig_1} show the DF timescale for both cases of a BBH with a total mass of 20 and 80 $\msun$.  For the lower-mass BBH, this timescale is shorter than the accretion timescale required to bring the component masses of the BBHs to resemble that of GW190521, while for the more massive BBH the timescales become comparable.

What mode of accretion would such a BBH experience when falling towards the dense inner regions of the halo?
The Eddington accretion rate for a BH, assuming 10\% radiative efficiency is given by:
\be
\dot{M}_{\rm Edd}=2.2\times10^{-8} \left(\frac{M}{M_{\odot}}\right) M_{\odot} \rm yr^{-1},
\ee
The radius inside which the BHL accretion rate exceeds the Eddington limit is marked by the dotted vertical lines in Figure \ref{fig_1}. Inside this region, radiative feedback makes accretion episodic and suppresses the time-averaged accretion rate below the Bondi rate, to a value near the Eddington limit~\citep{Milosavljevic+2009,ParkRicotti2012}.

Similarly, the radius at which the BHL accretion exceeds 500 times the Eddington limit is shown by vertical dot-dashed lines. 
This limit indicates where the BBH enters a phase of hyper-accretion, in which case photon trapping and rapid gas inflow suppress any negative radiative feedback, because the radiation from the accreting BH does not reach its gravitational influence radius. The Eddington limit, in this case, no longer applies~\citep{Inayoshi+2016,Sakurai+2016}. In such cases, only a few thousand years is required for a BBH with an initial mass of about 100 $\msun$ to accrete enough gas to resemble a system similar to GW190521.

\section{Predicted merger rate of GW190521-like binaries from Pop~III stars}

We next ask whether the extremely high redshift Pop~III star formation rate would be consistent with the merger rate for GW190521-like systems, which has been inferred from the LIGO observations to be  $0.13^{+0.3}_{-0.11} \rm Gpc^{-3} yr^{-1}$~\citep{Abbott:2020gz}. We adopt a SFR density (SFRD) from \citet{Visbal:2020bx} which 
has a value between $10^{-4}$ and $10^{-5}~{\rm M_\odot~yr^{-1}~Mpc^{-3}}$ at $z=10-20$, consistent with the constraint from the electron scattering optical depth $\tau_e$ measured by Planck~\citep{Visbal+2015}, including effects of Lyman-Werner radiation and setting a critical metallicity of
$3\times10^{-4}Z_{\odot}$ for Pop~III star formation.
We then convolve this SFRD with canonical delay time distributions \citep[DTD; see e.g.,][]{Safarzadeh:2019dn,Safarzadeh:2019dp,Safarzadeh:2019kj}:
\begin{align}
R(z)=&\int_{z_b=10}^{z_b=z} \lambda\frac{dP_m}{dt}(t-t_b-t_{\rm min})\psi(z_b)\frac{dt}{dz}(z_b)dz_b,
\end{align}
representing the time elapsed between the formation of the progenitor stars and the merger of the BBH,
where $dt/dz = -[(1+z) E(z) H_0]^{-1}$, and 
$E(z)=\sqrt{{\Omega}_{m,0}(1+z)^3+{\Omega}_{k,0}(1+z)^2+{\Omega}_{\Lambda}(z)}$. We use $H_{0} = 67~\mathrm{km\,s^{-1}\,Mpc^{-1}}$ for the Hubble constant, and $\Omega_{m,0} = 1-\Omega_{\Lambda} = 0.31$~\citep{Ade:2016xua}.
Here, $\lambda$ is the number of GW190521-like BBH systems per mass in Pop~III stars, which we set to $\lambda=10^{-4}\msun^{-1}$; 
$t_b$ is the cosmic time corresponding to redshift $z_b$; $dP_m/dt$ is the DTD, parametrized as a power-law distribution ($\propto t^{\Gamma}$) with a minimum delay time, $t_{\rm min}$, with a maximum merging timescale of 100 Gyr (the exact value of this maximum timescale does not affect our results). 
We note that the canonical slope of $\Gamma=-1$ comes from assuming the separation $a$ of binaries follows $dN/da\propto a^{-1}$, known as Opik's law.
We note that although Opik's law applies to the separation of massive stars, population studies of compact binary objects have shown $\Gamma=-1$ fits the merging timescale of the BBHs in the absence of gas dynamics physics explored in this work (i.e. assuming pure GW-driven inspiral; \citealt{Dominik2012}).  It is possible that gas dynamics driven inspiral \citep{Antoni2019} affects the shape of DTD distribution for the BBHs, which needs to be explored in  future work (see discussion below). 

\begin{figure}
\hspace{-0.2in}
\centering
\includegraphics[width=\columnwidth]{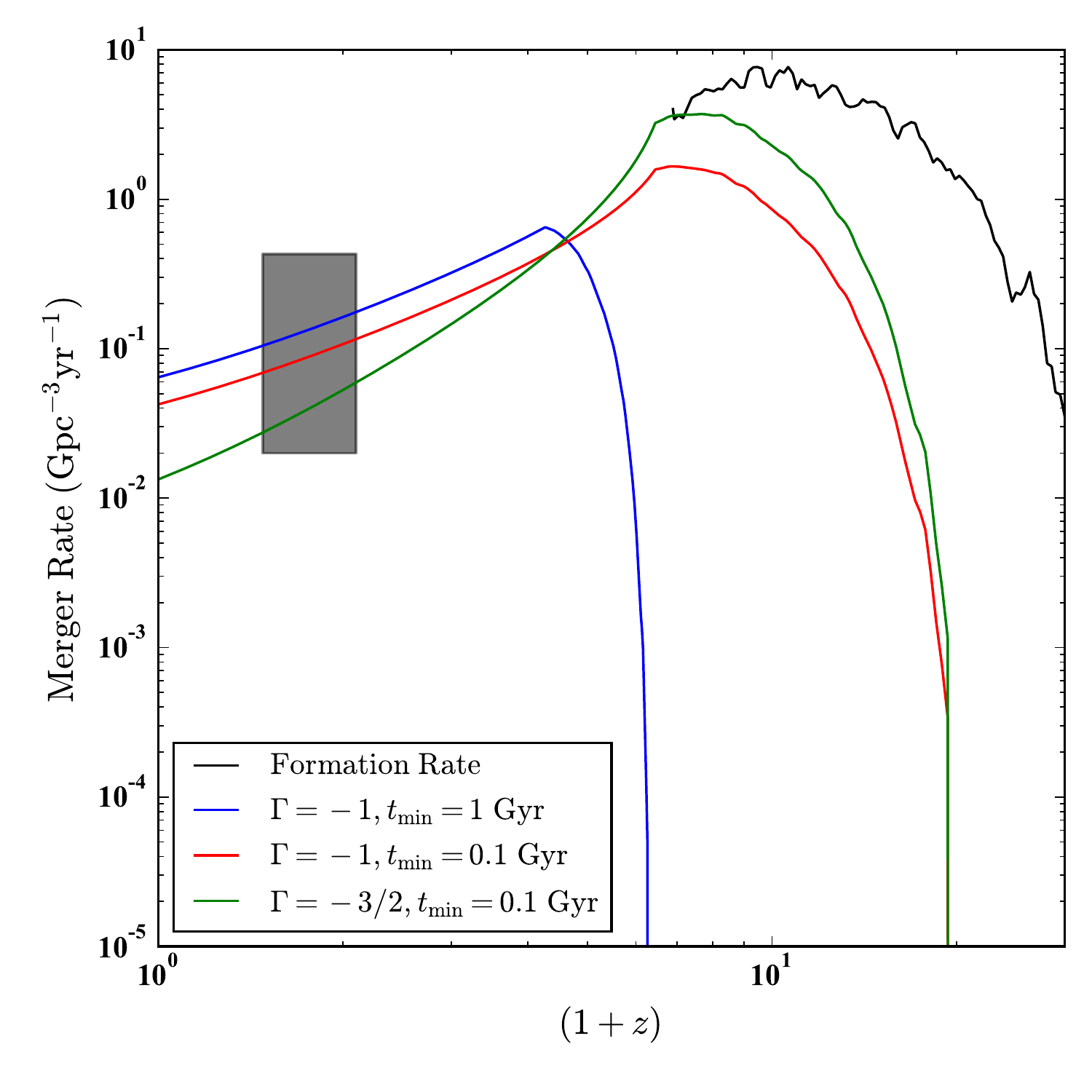}
\caption{The predicted merger rate of GW190521-like systems from the Pop~III SFRD \citep{Visbal:2020bx} assuming different delay-time distributions (DTDs) between formation and merger, and adopting a formation efficiency of $\lambda=10^{-4}\msun^{-1}$ for such binaries.
The grey shaded region is the redshift and merger rate estimates for GW190521 \citep{Abbott:2020dz}. Any canonical form of DTD can be adopted to explain the observed merger rate and redshift of GW190521.
}
\label{fig_2}
\end{figure}

The result is presented in Figure~\ref{fig_2}. With this choice of formation efficiency,  and given the large uncertainty in the inferred rate  (shown in Figure \ref{fig_2} by the shaded rectangle),
a wide range of canonical DTDs is consistent with the observed merger rate at the redshift of GW190521.  Note that the binaries merging at $z<1$ via GW-driven inspiral~\citep{Peters1964} had separations of $\sim 0.3~{\rm AU}\sim 10^{-6}$pc at $z>10$, much smaller than the BHL radius.

We also note that the adopted efficiency of $\lambda=10^{-4}\msun^{-1}$ is an upper limit since the Pop~III SFRD could be scaled up by nearly two orders of magnitude, depending on the choice of the IMF and the escape fraction of ionizing photons from massive stars, before the Planck $\tau_e$ constraint is violated~\citep{Visbal+2015,Inayoshi:2016du}.  As a result, it is sufficient for a fraction as small as a few percent of Pop~III remnants to undergo gas accretion and still be consistent with the lower end of the allowed range of rates inferred from GW190521. We note that such mergers will result in GW background with a spectral peak at a lower frequency \citep{Inayoshi:2016du}.

\section{Discussion}

Although here we have assumed a BHL mode of accretion, the accretion can be through a disk if there is a density/velocity gradient in the ambient medium or angular momentum is acquired in any other way.
In this case, a circumbinary accretion disk forms around the BBH, which will result in bringing the binary's mass ratio towards unity \citep[e.g., ][]{Farris+2014,ShiKrolik2015,DOrazio:2016gy,Duffell:2019vu}.
In their hydrodynamical simulations, \citet{Duffell:2019vu} find the following fitting formula for the ratio of accretion rates onto the individual binary components:
\be
\frac{\dot{m_2}}{\dot{m_1}}=\frac{1}{0.1+0.9q},
\ee
where $m_1$, and $m_2$ are the primary and the secondary component masses and $q\equiv m_2/m_1\leq 1$ is the mass ratio of the binary.
In this case, our results will be consistent with
$q=0.79^{+0.19}_{-0.29}$ inferred for GW190521~\citep{Abbott:2020gz} and the best-fitting value of $q=1$ found in a re-analysis of LIGO data \citet{Gayathri:2020ux} allowing for non-zero eccentricities of the binary orbit but using a sparser grid of template waveforms.

The reported effective spin of GW190521 indicates an in-plane effective spin of $\chi_p=0.68^{+0.25}_{-0.37}$, indicative of dynamical assembly. The error bars on the reported $\chi_p$ of this system are large and mostly coming from the merger and ringdown phase of the waveform. Moreover, the reported $\chi_p$ value stems from not being able to match the observed signal with other waveform templates and relying on $\chi_p$ as an extra degree of freedom to allow for a better fit to the observed signal. If this system is born as a remnant of a binary Pop~III stellar system, we might expect their spins to be aligned with the orbital AM 
vector~\citep{Bogdanovic+2007}; however, accretion onto the individual components through tilted "minidisks" may not be confined to the plane of the circumbinary disk~\citep{Nixon+2011}, and interactions with other flyby BHs can exert torques on a Pop~III BBH leading to spin misalignment with the orbital AM.

Our simple estimate of the accretion rate here is based on an illustrative toy model, mimicking the spherically-averaged density profiles found in simulations of atomic cooling halos. Three-dimensional simulations that have attempted to follow the trajectories of remnants BHs in early protogalaxies~\citep[e.g.][]{Alvarez+2009,Smith+2018,Pfister+2019} have found that they spend long periods in low-density regions away from the cores of the parent halos. We expect our binaries to have trajectories resembling low-eccentricity elliptical orbits such that they spend long enough times in the dense regions of the halo to increase their mass through accretion, similar to the behavior of single BHs that have been simulated so far in the literature. Nevertheless, our toy models suggest that it may be sufficient to spend tens of Myr in the inner $\sim$pc, where accretion can be near the Eddington limit, or as little as a few thousand years in the dense core where hyper-accretion occurs, for such 'wandering' BHs to $\sim$double their mass. The viability of this scenario needs to be confirmed in future, high-resolution simulations, which can accurately track the trajectories and accretion rates of individual stellar-mass BH remnants as they traverse the highly inhomogeneous inner regions of their parent halo, including the effects of gas drag and of radiative feedback on the nearby gas.

Finally, placing the binary in a gaseous environment, as envisioned here, can lead to gas-driven inspiral which is shorter than the inspiral time due to GW emission, and also shorter than the accretion timescale (although longer than the drag timescale on the center-of-mass motion). 
By simulating binaries in a common envelope (CE) phase \citet{Antoni2019} find that they merge before they can double their mass, for the ranges of parameters they investigated.  The regime in our scenario differs in important ways from those studied in the CE context. First, the binaries traversing the inhomogeneous gas distribution inside atomic cooling halos will ancounter gas with angular momentum, unlike the wind tunnel simulations of a CE phase. The resulting circumbinary disk can slow down (or even reverse) the binary inspiral. For example, \citet{Tiede2020} finds the ratio of the orbital evolution timescale and the accretion timescale  $-5 < (a/\dot{a})/(M_{\rm bh}/\dot{M_{\rm bh}})<3$ for circumbinary disks with Mach numbers in the range $10<\mathcal{M}<40$, implying that binaries in relatively warm disks can accrete significantly as they inspiral.  Second, in our case the binary separation is much smaller than its BHL radius, and gas drag from a BHL-like "wake" cannot operate. This regime has not been explored in CE simulations, and we expect the gas drag to become less important in such cases. Future work should address whether rapid inspiral in this regime prevents accretion, and/or modifies the DTDs.

While we have focused on the case of a BBH in a protogalaxy, we can alternatively consider single Pop~III remnant BHs that grow via accretion in the way described here. Cosmological N-body simulations have suggested that Pop~III remnant BHs are concentrated in the centers of halos and subhalos~\citep[e.g.][]{Ishiyama+2016}, and the heaviest remnants would sink to the innermost regions~\citep{MadauRees2001} and participate in dynamical capture processes at lower redshifts. While a merger of two such Pop~III remnants is likely rare, in the case of GW190521, an alternative possibility is that only the primary is above the PI limit (see related discussion in \citealt{Fishbach2020}). This could be explained as the dynamical pairing of a Pop~III remnant BH with a less massive companion, the latter having formed as a remnant of normal massive star formation at lower redshift.

Finally, if either single or binary Pop~III BHs can grow via accretion, then  even more massive BBHs, possibly well above that of GW190521, may be expected through this channel. The detectability of these events will be suppressed as they redshift out of the sensitivity band of the current ground-based detectors. However, observations with future instruments covering lower frequencies, such as DECIGO~\citep{DECIGO} or LISA~\citep{LISA}, can shed light on the prevalence of the  mechanism proposed here.

\acknowledgements 
We are thankful to the referee for their constructive comments, to Eli Visbal for providing us the data for the SFRD of Pop~III stars, and to Ilya Mandel, Stan Woosley, Kohei Inayoshi, Enrico Ramirez-Ruiz, Martin Rees and Jerry Ostriker for insightful discussions. 
MTS thanks the Heising-Simons Foundation, the Danish National Research Foundation (DNRF132), and NSF (AST-1911206 and AST-1852393) and
ZH thanks NSF grants AST-2006176 and 1715661, and NASA grant NNX17AL82G for support.

\bibliographystyle{yahapj}
\bibliography{the_entire_lib}

\begin{thebibliography}{}
\providecommand\natexlab[1]{#1}
\providecommand\JournalTitle[1]{#1}

\bibitem[{Abbott {et~al.}(2020{\natexlab{a}})Abbott, Abbott, Abraham, Acernese,
  Ackley, Adams, Adhikari, Adya, Affeldt, Agathos, Agatsuma, Aggarwal, Aguiar,
  Aich, Aiello, Ain, Ajith, Akcay, Allen, Allocca, Altin, Amato, Anand,
  Ananyeva, Anderson, Anderson, Angelova, Ansoldi, Antier, Appert, Arai, Araya,
  Areeda, Ar{\`e}ne, Arnaud, Aronson, Arun, Asali, Ascenzi, Ashton, Aston,
  Astone, Aubin, Aufmuth, AultONeal, Austin, Avendano, Babak, Bacon, Badaracco,
  Bader, Bae, Baer, Baird, Baldaccini, Ballardin, Ballmer, Bals, Balsamo,
  Baltus, Banagiri, Bankar, Bankar, Barayoga, Barbieri, Barish, Barker,
  Barkett, Barneo, Barone, Barr, Barsotti, Barsuglia, Barta, Bartlett, Bartos,
  Bassiri, Basti, Bawaj, Bayley, Bazzan, B{\'e}csy, Bejger, Belahcene, Bell,
  Beniwal, Benjamin, Bentley, Bergamin, Berger, Bergmann, Bernuzzi, Berry,
  Bersanetti, Bertolini, Betzwieser, Bhandare, Bhandari, Bidler, Biggs,
  Bilenko, Billingsley, Birney, Birnholtz, Biscans, Bischi, Biscoveanu, Bisht,
  Bissenbayeva, Bitossi, Bizouard, Blackburn, Blackman, Blair, Blair, Blair,
  Bobba, Bode, Boer, Boetzel, Bogaert, Bondu, Bonilla, Bonnand, Booker, Boom,
  Bork, Boschi, Bose, Bossilkov, Bosveld, Bouffanais, Bozzi, Bradaschia, Brady,
  Bramley, Branchesi, Brau, Breschi, Briant, Briggs, Brighenti, Brillet,
  Brinkmann, Brockill, Brooks, Brooks, Brown, Brunett, Bruno, Bruntz, Buikema,
  Bulik, Bulten, Buonanno, Buscicchio, Buskulic, Byer, Cabero, Cadonati,
  Cagnoli, Cahillane, Calder{\'o}n~Bustillo, Callaghan, Callister, Calloni,
  Camp, Canepa, Cannon, Cao, Cao, Carapella, Carbognani, Caride, Carney,
  Carullo, Casanueva~Diaz, Casentini, Casta{\~n}eda, Caudill, Cavagli{\`a},
  Cavalier, Cavalieri, Cella, Cerd{\'a}-Dur{\'a}n, Cesarini, Chaibi,
  Chakravarti, Chan, Chan, Chandra, Chao, Charlton, Chase, Chassande-Mottin,
  Chatterjee, Chaturvedi, Chatziioannou, Chen, Chen, Chen, Cheng, Cheong, Chia,
  Chiadini, Chierici, Chincarini, Chiummo, Cho, Cho, Cho, Christensen, Chu,
  Chua, Chung, Chung, Ciani, Ciecielag, Cie{\'{s}}lar, Ciobanu, Ciolfi,
  Cipriano, Cirone, Clara, Clark, Clearwater, Clesse, Cleva, Coccia, Cohadon,
  Cohen, Colleoni, Collette, Collins, Colpi, Constancio, Conti, Cooper, Corban,
  Corbitt, Cordero-Carri{\'o}n, Corezzi, Corley, Cornish, Corre, Corsi,
  Cortese, Costa, Cotesta, Coughlin, Coughlin, Coulon, Countryman, Couvares,
  Covas, Coward, Cowart, Coyne, Coyne, Creighton, Creighton, Cripe, Croquette,
  Crowder, Cudell, Cullen, Cumming, Cummings, Cunningham, Cuoco, Curylo,
  Canton, D{\'a}lya, Dana, Daneshgaran-Bajastani, D'Angelo, Danilishin,
  D'Antonio, Danzmann, Darsow-Fromm, Dasgupta, Datrier, Dattilo, Dave, Davier,
  Davies, Davis, Daw, DeBra, Deenadayalan, Degallaix, De~Laurentis,
  Del{\'e}glise, Delfavero, De~Lillo, Del~Pozzo, DeMarchi, D'Emilio, Demos,
  Dent, De~Pietri, De~Rosa, De~Rossi, DeSalvo, de~Varona, \&
  Dhur...}]{Abbott:2020dz}
Abbott, R., Abbott, T.~D., Abraham, S., {et~al.} 2020{\natexlab{a}},
  \href{http://dx.doi.org/10.1103/PhysRevLett.125.101102}{\JournalTitle{\prl},
  125, 101102}

\bibitem[{Abbott {et~al.}(2020{\natexlab{b}})Abbott, Abbott, Abraham, Acernese,
  Ackley, Adams, Adhikari, Adya, Affeldt, Agathos, Agatsuma, Aggarwal, Aguiar,
  Aich, Aiello, Ain, Ajith, Akcay, Allen, Allocca, Altin, Amato, Anand,
  Ananyeva, Anderson, Anderson, Angelova, Ansoldi, Antier, Appert, Arai, Araya,
  Areeda, Ar{\`e}ne, Arnaud, Aronson, Arun, Asali, Ascenzi, Ashton, Aston,
  Astone, Aubin, Aufmuth, AultONeal, Austin, Avendano, Babak, Bacon, Badaracco,
  Bader, Bae, Baer, Baird, Baldaccini, Ballardin, Ballmer, Bals, Balsamo,
  Baltus, Banagiri, Bankar, Bankar, Barayoga, Barbieri, Barish, Barker,
  Barkett, Barneo, Barone, Barr, Barsotti, Barsuglia, Barta, Bartlett, Bartos,
  Bassiri, Basti, Bawaj, Bayley, Bazzan, B{\'e}csy, Bejger, Belahcene, Bell,
  Beniwal, Benjamin, Bentley, Bergamin, Berger, Bergmann, Bernuzzi, Berry,
  Bersanetti, Bertolini, Betzwieser, Bhandare, Bhandari, Bidler, Biggs,
  Bilenko, Billingsley, Birney, Birnholtz, Biscans, Bischi, Biscoveanu, Bisht,
  Bissenbayeva, Bitossi, Bizouard, Blackburn, Blackman, Blair, Blair, Blair,
  Bobba, Bode, Boer, Boetzel, Bogaert, Bondu, Bonilla, Bonnand, Booker, Boom,
  Bork, Boschi, Bose, Bossilkov, Bosveld, Bouffanais, Bozzi, Bradaschia, Brady,
  Bramley, Branchesi, Brau, Breschi, Briant, Briggs, Brighenti, Brillet,
  Brinkmann, Brockill, Brooks, Brooks, Brown, Brunett, Bruno, Bruntz, Buikema,
  Bulik, Bulten, Buonanno, Buscicchio, Buskulic, Byer, Cabero, Cadonati,
  Cagnoli, Cahillane, Bustillo, Callaghan, Callister, Calloni, Camp, Canepa,
  Cannon, Cao, Cao, Carapella, Carbognani, Caride, Carney, Carullo, Diaz,
  Casentini, Casta{\~n}eda, Caudill, Cavagli{\`a}, Cavalier, Cavalieri, Cella,
  Cerd{\'a}-Dur{\'a}n, Cesarini, Chaibi, Chakravarti, Chan, Chan, Chao,
  Charlton, Chase, Chassande-Mottin, Chatterjee, Chaturvedi, Chatziioannou,
  Chen, Chen, Chen, Cheng, Cheong, Chia, Chiadini, Chierici, Chincarini,
  Chiummo, Cho, Cho, Cho, Christensen, Chu, Chua, Chung, Chung, Ciani,
  Ciecielag, Cie{\'{s}}lar, Ciobanu, Ciolfi, Cipriano, Cirone, Clara, Clark,
  Clearwater, Clesse, Cleva, Coccia, Cohadon, Cohen, Colleoni, Collette,
  Collins, Colpi, Constancio, Conti, Cooper, Corban, Corbitt,
  Cordero-Carri{\'o}n, Corezzi, Corley, Cornish, Corre, Corsi, Cortese, Costa,
  Cotesta, Coughlin, Coughlin, Coulon, Countryman, Couvares, Covas, Coward,
  Cowart, Coyne, Coyne, Creighton, Creighton, Cripe, Croquette, Crowder,
  Cudell, Cullen, Cumming, Cummings, Cunningham, Cuoco, Curylo, Canton,
  D{\'a}lya, Dana, Daneshgaran-Bajastani, D'Angelo, Danilishin, D'Antonio,
  Danzmann, Darsow-Fromm, Dasgupta, Datrier, Dattilo, Dave, Davier, Davies,
  Davis, Daw, DeBra, Deenadayalan, Degallaix, De~Laurentis, Del{\'e}glise,
  Delfavero, De~Lillo, Del~Pozzo, DeMarchi, D'Emilio, Demos, Dent, De~Pietri,
  De~Rosa, De~Rossi, DeSalvo, de~Varona, \& Dhurandhar}]{Abbott:2020gz}
---. 2020{\natexlab{b}},
  \href{http://dx.doi.org/10.3847/2041-8213/aba493}{\JournalTitle{\apjl}, 900,
  L13}

\bibitem[{{Alvarez} {et~al.}(2009){Alvarez}, {Wise}, \& {Abel}}]{Alvarez+2009}
{Alvarez}, M.~A., {Wise}, J.~H., \& {Abel}, T. 2009,
  \href{http://dx.doi.org/10.1088/0004-637X/701/2/L133}{\JournalTitle{\apjl},
  701, L133}

\bibitem[{{Amaro-Seoane} {et~al.}(2017){Amaro-Seoane}, {Audley}, {Babak},
  {Baker}, {Barausse}, {Bender}, {Berti}, {Binetruy}, {Born}, {Bortoluzzi},
  {Camp}, {Caprini}, {Cardoso}, {Colpi}, {Conklin}, {Cornish}, {Cutler},
  {Danzmann}, {Dolesi}, {Ferraioli}, {Ferroni}, {Fitzsimons}, {Gair}, {Gesa
  Bote}, {Giardini}, {Gibert}, {Grimani}, {Halloin}, {Heinzel}, {Hertog},
  {Hewitson}, {Holley-Bockelmann}, {Hollington}, {Hueller}, {Inchauspe},
  {Jetzer}, {Karnesis}, {Killow}, {Klein}, {Klipstein}, {Korsakova}, {Larson},
  {Livas}, {Lloro}, {Man}, {Mance}, {Martino}, {Mateos}, {McKenzie},
  {McWilliams}, {Miller}, {Mueller}, {Nardini}, {Nelemans}, {Nofrarias},
  {Petiteau}, {Pivato}, {Plagnol}, {Porter}, {Reiche}, {Robertson},
  {Robertson}, {Rossi}, {Russano}, {Schutz}, {Sesana}, {Shoemaker}, {Slutsky},
  {Sopuerta}, {Sumner}, {Tamanini}, {Thorpe}, {Troebs}, {Vallisneri},
  {Vecchio}, {Vetrugno}, {Vitale}, {Volonteri}, {Wanner}, {Ward}, {Wass},
  {Weber}, {Ziemer}, \& {Zweifel}}]{LISA}
{Amaro-Seoane}, P., {Audley}, H., {Babak}, S., {et~al.} 2017,
  \JournalTitle{Proposal submitted to ESA on January 13th, 2007; e-print
  arXiv:1702.00786}

\bibitem[{{Antoni} {et~al.}(2019){Antoni}, {MacLeod}, \&
  {Ramirez-Ruiz}}]{Antoni2019}
{Antoni}, A., {MacLeod}, M., \& {Ramirez-Ruiz}, E. 2019,
  \href{http://dx.doi.org/10.3847/1538-4357/ab3466}{\JournalTitle{\apj}, 884,
  22}

\bibitem[{{Bogdanovi{\'c}} {et~al.}(2007){Bogdanovi{\'c}}, {Reynolds}, \&
  {Miller}}]{Bogdanovic+2007}
{Bogdanovi{\'c}}, T., {Reynolds}, C.~S., \& {Miller}, M.~C. 2007,
  \href{http://dx.doi.org/10.1086/518769}{\JournalTitle{\apjl}, 661, L147}

\bibitem[{{Bondi}(1952)}]{Bondi:1952fc}
{Bondi}, H. 1952,
  \href{http://dx.doi.org/10.1093/mnras/112.2.195}{\JournalTitle{\mnras}, 112,
  195}

\bibitem[{{Bondi} \& {Hoyle}(1944)}]{Bondi:1944gc}
{Bondi}, H., \& {Hoyle}, F. 1944,
  \href{http://dx.doi.org/10.1093/mnras/104.5.273}{\JournalTitle{\mnras}, 104,
  273}

\bibitem[{{Chon} \& {Hosokawa}(2019)}]{Chon+2019}
{Chon}, S., \& {Hosokawa}, T. 2019,
  \href{http://dx.doi.org/10.1093/mnras/stz1824}{\JournalTitle{\mnras}, 488,
  2658}

\bibitem[{{Clark} {et~al.}(2011){Clark}, {Glover}, {Klessen}, \&
  {Bromm}}]{Clark2011}
{Clark}, P.~C., {Glover}, S. C.~O., {Klessen}, R.~S., \& {Bromm}, V. 2011,
  \href{http://dx.doi.org/10.1088/0004-637X/727/2/110}{\JournalTitle{\apj},
  727, 110}

\bibitem[{{De Luca} {et~al.}(2020){De Luca}, {Desjacques}, {Franciolini},
  {Pani}, \& {Riotto}}]{Luca2020}
{De Luca}, V., {Desjacques}, V., {Franciolini}, G., {Pani}, P., \& {Riotto}, A.
  2020, \JournalTitle{arXiv e-prints}, arXiv:2009.01728

\bibitem[{{Dominik} {et~al.}(2012){Dominik}, {Belczynski}, {Fryer}, {Holz},
  {Berti}, {Bulik}, {Mand el}, \& {O'Shaughnessy}}]{Dominik2012}
{Dominik}, M., {Belczynski}, K., {Fryer}, C., {et~al.} 2012,
  \href{http://dx.doi.org/10.1088/0004-637X/759/1/52}{\JournalTitle{\apj}, 759,
  52}

\bibitem[{{D'Orazio} {et~al.}(2016){D'Orazio}, {Haiman}, {Duffell},
  {MacFadyen}, \& {Farris}}]{DOrazio:2016gy}
{D'Orazio}, D.~J., {Haiman}, Z., {Duffell}, P., {MacFadyen}, A., \& {Farris},
  B. 2016, \href{http://dx.doi.org/10.1093/mnras/stw792}{\JournalTitle{\mnras},
  459, 2379}

\bibitem[{Duffell {et~al.}(2019)Duffell, D'Orazio, Derdzinski, Haiman,
  MacFadyen, Rosen, \& Zrake}]{Duffell:2019vu}
Duffell, P.~C., D'Orazio, D., Derdzinski, A., {et~al.} 2019,
  \JournalTitle{arXiv}, \href{http://arxiv.org/abs/1911.05506}{{\sffamily
  1911.05506}}

\bibitem[{{Farmer} {et~al.}(2020){Farmer}, {Renzo}, {de Mink}, {Fishbach}, \&
  {Justham}}]{Farmer2020}
{Farmer}, R., {Renzo}, M., {de Mink}, S., {Fishbach}, M., \& {Justham}, S.
  2020, \JournalTitle{arXiv e-prints}, arXiv:2006.06678

\bibitem[{{Farrell} {et~al.}(2020){Farrell}, {Groh}, {Hirschi}, {Murphy},
  {Kaiser}, {Ekstr{\"o}m}, {Georgy}, \& {Meynet}}]{Farrell+2020}
{Farrell}, E.~J., {Groh}, J.~H., {Hirschi}, R., {et~al.} 2020,
  \JournalTitle{\mnras, submitted; e-print arXiv:2009.06585}

\bibitem[{{Farris} {et~al.}(2014){Farris}, {Duffell}, {MacFadyen}, \&
  {Haiman}}]{Farris+2014}
{Farris}, B.~D., {Duffell}, P., {MacFadyen}, A.~I., \& {Haiman}, Z. 2014,
  \href{http://dx.doi.org/10.1088/0004-637X/783/2/134}{\JournalTitle{\apj},
  783, 134}

\bibitem[{{Farris} {et~al.}(2010){Farris}, {Liu}, \& {Shapiro}}]{Farris+2010}
{Farris}, B.~D., {Liu}, Y.~T., \& {Shapiro}, S.~L. 2010,
  \href{http://dx.doi.org/10.1103/PhysRevD.81.084008}{\JournalTitle{\prd}, 81,
  084008}

\bibitem[{{Fishbach} \& {Holz}(2020)}]{Fishbach2020}
{Fishbach}, M., \& {Holz}, D.~E. 2020, \JournalTitle{arXiv e-prints},
  arXiv:2009.05472

\bibitem[{Fragione {et~al.}(2020)Fragione, Loeb, \& Rasio}]{Fragione:2020uh}
Fragione, G., Loeb, A., \& Rasio, F.~A. 2020, \JournalTitle{arXiv},
  \href{http://arxiv.org/abs/2009.05065}{{\sffamily 2009.05065}}

\bibitem[{Gayathri {et~al.}(2020)Gayathri, Healy, Lange, O'Brien, Szczepanczyk,
  Bartos, Campanelli, Klimenko, Lousto, \&
  O{\textquoteright}Shaughnessy}]{Gayathri:2020ux}
Gayathri, V., Healy, J., Lange, J., {et~al.} 2020, \JournalTitle{arXiv},
  \href{http://arxiv.org/abs/2009.05461}{{\sffamily 2009.05461}}

\bibitem[{{Hirano} {et~al.}(2015){Hirano}, {Hosokawa}, {Yoshida}, {Omukai}, \&
  {Yorke}}]{Hirano+2015}
{Hirano}, S., {Hosokawa}, T., {Yoshida}, N., {Omukai}, K., \& {Yorke}, H.~W.
  2015, \href{http://dx.doi.org/10.1093/mnras/stv044}{\JournalTitle{\mnras},
  448, 568}

\bibitem[{{Hirano} {et~al.}(2018){Hirano}, {Yoshida}, {Sakurai}, \&
  {Fujii}}]{Hirano2018}
{Hirano}, S., {Yoshida}, N., {Sakurai}, Y., \& {Fujii}, M.~S. 2018,
  \href{http://dx.doi.org/10.3847/1538-4357/aaaaba}{\JournalTitle{\apj}, 855,
  17}

\bibitem[{{Hoyle} \& {Lyttleton}(1940)}]{Hoyle:1939fl}
{Hoyle}, F., \& {Lyttleton}, R.~A. 1940,
  \href{http://dx.doi.org/10.1017/S0305004100017369}{\JournalTitle{Proceedings
  of the Cambridge Philosophical Society}, 36, 325}

\bibitem[{{Inayoshi} {et~al.}(2016{\natexlab{a}}){Inayoshi}, {Haiman}, \&
  {Ostriker}}]{Inayoshi+2016}
{Inayoshi}, K., {Haiman}, Z., \& {Ostriker}, J.~P. 2016{\natexlab{a}},
  \href{http://dx.doi.org/10.1093/mnras/stw836}{\JournalTitle{\mnras}, 459,
  3738}

\bibitem[{{Inayoshi} {et~al.}(2016{\natexlab{b}}){Inayoshi}, {Kashiyama},
  {Visbal}, \& {Haiman}}]{Inayoshi:2016du}
{Inayoshi}, K., {Kashiyama}, K., {Visbal}, E., \& {Haiman}, Z.
  2016{\natexlab{b}},
  \href{http://dx.doi.org/10.1093/mnras/stw1431}{\JournalTitle{\mnras}, 461,
  2722}

\bibitem[{{Inayoshi} {et~al.}(2019){Inayoshi}, {Visbal}, \&
  {Haiman}}]{Inayoshi+2020}
{Inayoshi}, K., {Visbal}, E., \& {Haiman}, Z. 2019, \JournalTitle{\araa, in
  press; e-print arXiv:1911.05791}

\bibitem[{{Ishiyama} {et~al.}(2016){Ishiyama}, {Sudo}, {Yokoi}, {Hasegawa},
  {Tominaga}, \& {Susa}}]{Ishiyama+2016}
{Ishiyama}, T., {Sudo}, K., {Yokoi}, S., {et~al.} 2016,
  \href{http://dx.doi.org/10.3847/0004-637X/826/1/9}{\JournalTitle{\apj}, 826,
  9}

\bibitem[{{Johnson} \& {Bromm}(2007)}]{JohnsonBromm2007}
{Johnson}, J.~L., \& {Bromm}, V. 2007,
  \href{http://dx.doi.org/10.1111/j.1365-2966.2006.11275.x}{\JournalTitle{\mnras},
  374, 1557}

\bibitem[{{Kawamura} {et~al.}(2011){Kawamura}, {Ando}, {Seto}, {Sato},
  {Nakamura}, {Tsubono}, {Kand a}, {Tanaka}, {Yokoyama}, {Funaki}, {Numata},
  {Ioka}, {Takashima}, {Agatsuma}, {Akutsu}, {Aoyanagi}, {Arai}, {Araya},
  {Asada}, {Aso}, {Chen}, {Chiba}, {Ebisuzaki}, {Ejiri}, {Enoki}, {Eriguchi},
  {Fujimoto}, {Fujita}, {Fukushima}, {Futamase}, {Harada}, {Hashimoto},
  {Hayama}, {Hikida}, {Himemoto}, {Hirabayashi}, {Hiramatsu}, {Hong},
  {Horisawa}, {Hosokawa}, {Ichiki}, {Ikegami}, {Inoue}, {Ishidoshiro},
  {Ishihara}, {Ishikawa}, {Ishizaki}, {Ito}, {Itoh}, {Izumi}, {Kawano},
  {Kawashima}, {Kawazoe}, {Kishimoto}, {Kiuchi}, {Kobayashi}, {Kohri},
  {Koizumi}, {Kojima}, {Kokeyama}, {Kokuyama}, {Kotake}, {Kozai}, {Kunimori},
  {Kuninaka}, {Kuroda}, {Kuroyanagi}, {Maeda}, {Matsuhara}, {Matsumoto},
  {Michimura}, {Miyakawa}, {Miyamoto}, {Miyoki}, {Morimoto}, {Morisawa},
  {Moriwaki}, {Mukohyama}, {Musha}, {Nagano}, {Naito}, {Nakamura}, {Nakano},
  {Nakao}, {Nakasuka}, {Nakayama}, {Nakazawa}, {Nishida}, {Nishiyama},
  {Nishizawa}, {Niwa}, {Noumi}, {Obuchi}, {Ohashi}, {Ohishi}, {Ohkawa},
  {Okada}, {Okada}, {Oohara}, {Sago}, {Saijo}, {Saito}, {Sakagami}, {Sakai},
  {Sakata}, {Sasaki}, {Sato}, {Shibata}, {Shinkai}, {Shoda}, {Somiya},
  {Sotani}, {Sugiyama}, {Suwa}, {Suzuki}, {Tagoshi}, {Takahashi}, {Takahashi},
  {Takahashi}, {Takahashi}, {Takahashi}, {Takahashi}, {Takahashi}, {Akiteru},
  {Takano}, {Tanaka}, {Taniguchi}, {Taruya}, {Tashiro}, {Torii}, {Toyoshima},
  {Tsujikawa}, {Tsunesada}, {Ueda}, {Ueda}, {Utashima}, {Wakabayashi}, {Yagi},
  {Yamakawa}, {Yamamoto}, {Yamazaki}, {Yoo}, {Yoshida}, {Yoshino}, \&
  {Sun}}]{DECIGO}
{Kawamura}, S., {Ando}, M., {Seto}, N., {et~al.} 2011,
  \href{http://dx.doi.org/10.1088/0264-9381/28/9/094011}{\JournalTitle{Classical
  and Quantum Gravity}, 28, 094011}

\bibitem[{{Kinugawa} {et~al.}(2014){Kinugawa}, {Inayoshi}, {Hotokezaka},
  {Nakauchi}, \& {Nakamura}}]{Kinugawa+2014}
{Kinugawa}, T., {Inayoshi}, K., {Hotokezaka}, K., {Nakauchi}, D., \&
  {Nakamura}, T. 2014,
  \href{http://dx.doi.org/10.1093/mnras/stu1022}{\JournalTitle{\mnras}, 442,
  2963}

\bibitem[{{Kinugawa} {et~al.}(2020){Kinugawa}, {Nakamura}, \&
  {Nakano}}]{Kinugawa+2020}
{Kinugawa}, T., {Nakamura}, T., \& {Nakano}, H. 2020, \JournalTitle{\mnras,
  submitted; e-print arXiv:2009.06922}

\bibitem[{{Kitayama} {et~al.}(2004){Kitayama}, {Yoshida}, {Susa}, \&
  {Umemura}}]{Kitayama+2004}
{Kitayama}, T., {Yoshida}, N., {Susa}, H., \& {Umemura}, M. 2004,
  \href{http://dx.doi.org/10.1086/423313}{\JournalTitle{\apj}, 613, 631}

\bibitem[{{Madau} \& {Rees}(2001)}]{MadauRees2001}
{Madau}, P., \& {Rees}, M.~J. 2001,
  \href{http://dx.doi.org/10.1086/319848}{\JournalTitle{\apjl}, 551, L27}

\bibitem[{{McKernan} {et~al.}(2012){McKernan}, {Ford}, {Lyra}, \&
  {Perets}}]{McKernan+2012}
{McKernan}, B., {Ford}, K.~E.~S., {Lyra}, W., \& {Perets}, H.~B. 2012,
  \href{http://dx.doi.org/10.1111/j.1365-2966.2012.21486.x}{\JournalTitle{\mnras},
  425, 460}

\bibitem[{{Milosavljevi{\'c}} {et~al.}(2009){Milosavljevi{\'c}}, {Couch}, \&
  {Bromm}}]{Milosavljevic+2009}
{Milosavljevi{\'c}}, M., {Couch}, S.~M., \& {Bromm}, V. 2009,
  \href{http://dx.doi.org/10.1088/0004-637X/696/2/L146}{\JournalTitle{\apjl},
  696, L146}

\bibitem[{{Nakamura} \& {Umemura}(2001)}]{Nakamura2001}
{Nakamura}, F., \& {Umemura}, M. 2001,
  \href{http://dx.doi.org/10.1086/318663}{\JournalTitle{\apj}, 548, 19}

\bibitem[{{Nixon} {et~al.}(2011){Nixon}, {Cossins}, {King}, \&
  {Pringle}}]{Nixon+2011}
{Nixon}, C.~J., {Cossins}, P.~J., {King}, A.~R., \& {Pringle}, J.~E. 2011,
  \href{http://dx.doi.org/10.1111/j.1365-2966.2010.17952.x}{\JournalTitle{\mnras},
  412, 1591}

\bibitem[{{Park} \& {Ricotti}(2012)}]{ParkRicotti2012}
{Park}, K., \& {Ricotti}, M. 2012,
  \href{http://dx.doi.org/10.1088/0004-637X/747/1/9}{\JournalTitle{\apj}, 747,
  9}

\bibitem[{{Perna} {et~al.}(2019){Perna}, {Wang}, {Farr}, {Leigh}, \&
  {Cantiello}}]{Perna2019}
{Perna}, R., {Wang}, Y.-H., {Farr}, W.~M., {Leigh}, N., \& {Cantiello}, M.
  2019, \href{http://dx.doi.org/10.3847/2041-8213/ab2336}{\JournalTitle{\apjl},
  878, L1}

\bibitem[{{Peters}(1964)}]{Peters1964}
{Peters}, P.~C. 1964,
  \href{http://dx.doi.org/10.1103/PhysRev.136.B1224}{\JournalTitle{Physical
  Review}, 136, 1224}

\bibitem[{{Pfister} {et~al.}(2019){Pfister}, {Volonteri}, {Dubois}, {Dotti}, \&
  {Colpi}}]{Pfister+2019}
{Pfister}, H., {Volonteri}, M., {Dubois}, Y., {Dotti}, M., \& {Colpi}, M. 2019,
  \href{http://dx.doi.org/10.1093/mnras/stz822}{\JournalTitle{\mnras}, 486,
  101}

\bibitem[{{Planck Collaboration} {et~al.}(2016){Planck Collaboration}, {Ade},
  {Aghanim}, {Arnaud}, {Ashdown}, {Aumont}, {Baccigalupi}, {Banday},
  {Barreiro}, {Bartlett}, {Bartolo}, {Battaner}, {Battye}, {Benabed},
  {Beno{\^\i}t}, {Benoit-L{\'e}vy}, {Bernard}, {Bersanelli}, {Bielewicz},
  {Bock}, {Bonaldi}, {Bonavera}, {Bond}, {Borrill}, {Bouchet}, {Boulanger},
  {Bucher}, {Burigana}, {Butler}, {Calabrese}, {Cardoso}, {Catalano},
  {Challinor}, {Chamballu}, {Chary}, {Chiang}, {Chluba}, {Christensen},
  {Church}, {Clements}, {Colombi}, {Colombo}, {Combet}, {Coulais}, {Crill},
  {Curto}, {Cuttaia}, {Danese}, {Davies}, {Davis}, {de Bernardis}, {de Rosa},
  {de Zotti}, {Delabrouille}, {D{\'e}sert}, {Di Valentino}, {Dickinson},
  {Diego}, {Dolag}, {Dole}, {Donzelli}, {Dor{\'e}}, {Douspis}, {Ducout},
  {Dunkley}, {Dupac}, {Efstathiou}, {Elsner}, {En{\ss}lin}, {Eriksen},
  {Farhang}, {Fergusson}, {Finelli}, {Forni}, {Frailis}, {Fraisse},
  {Franceschi}, {Frejsel}, {Galeotta}, {Galli}, {Ganga}, {Gauthier}, {Gerbino},
  {Ghosh}, {Giard}, {Giraud-H{\'e}raud}, {Giusarma}, {Gjerl{\o}w},
  {Gonz{\'a}lez-Nuevo}, {G{\'o}rski}, {Gratton}, {Gregorio}, {Gruppuso},
  {Gudmundsson}, {Hamann}, {Hansen}, {Hanson}, {Harrison}, {Helou},
  {Henrot-Versill{\'e}}, {Hern{\'a}ndez-Monteagudo}, {Herranz}, {Hildebrand t},
  {Hivon}, {Hobson}, {Holmes}, {Hornstrup}, {Hovest}, {Huang}, {Huffenberger},
  {Hurier}, {Jaffe}, {Jaffe}, {Jones}, {Juvela}, {Keih{\"a}nen}, {Keskitalo},
  {Kisner}, {Kneissl}, {Knoche}, {Knox}, {Kunz}, {Kurki-Suonio}, {Lagache},
  {L{\"a}hteenm{\"a}ki}, {Lamarre}, {Lasenby}, {Lattanzi}, {Lawrence}, {Leahy},
  {Leonardi}, {Lesgourgues}, {Levrier}, {Lewis}, {Liguori}, {Lilje},
  {Linden-V{\o}rnle}, {L{\'o}pez-Caniego}, {Lubin}, {Mac{\'\i}as-P{\'e}rez},
  {Maggio}, {Maino}, {Mandolesi}, {Mangilli}, {Marchini}, {Maris}, {Martin},
  {Martinelli}, {Mart{\'\i}nez-Gonz{\'a}lez}, {Masi}, {Matarrese}, {McGehee},
  {Meinhold}, {Melchiorri}, {Melin}, {Mendes}, {Mennella}, {Migliaccio},
  {Millea}, {Mitra}, {Miville-Desch{\^e}nes}, {Moneti}, {Montier}, {Morgante},
  {Mortlock}, {Moss}, {Munshi}, {Murphy}, {Naselsky}, {Nati}, {Natoli},
  {Netterfield}, {N{\o}rgaard-Nielsen}, {Noviello}, {Novikov}, {Novikov},
  {Oxborrow}, {Paci}, {Pagano}, {Pajot}, {Paladini}, {Paoletti}, {Partridge},
  {Pasian}, {Patanchon}, {Pearson}, {Perdereau}, {Perotto}, {Perrotta},
  {Pettorino}, {Piacentini}, {Piat}, {Pierpaoli}, {Pietrobon}, {Plaszczynski},
  {Pointecouteau}, {Polenta}, {Popa}, {Pratt}, {Pr{\'e}zeau}, {Prunet},
  {Puget}, {Rachen}, {Reach}, {Rebolo}, {Reinecke}, {Remazeilles}, {Renault},
  {Renzi}, {Ristorcelli}, {Rocha}, {Rosset}, {Rossetti}, {Roudier},
  {Rouill{\'e} d'Orfeuil}, {Rowan-Robinson}, {Rubi{\~n}o-Mart{\'\i}n},
  {Rusholme}, {Said}, {Salvatelli}, {Salvati}, {Sandri}, {Santos},
  {Savelainen}, {Savini}, {Scott}, {Seiffert}, {Serra}, {Shellard}, {Spencer},
  {Spinelli}, {Stolyarov}, {Stompor}, {Sudiwala}, {Sunyaev}, {Sutton},
  {Suur-Uski}, {Sygnet}, {Tauber}, {Terenzi}, {Toffolatti}, {Tomasi},
  {Tristram}, {Trombetti}, {Tucci}, {Tuovinen}, {T{\"u}rler}, {Umana},
  {Valenziano}, {Valiviita}, {Van Tent}, {Vielva}, {Villa}, {Wade}, {Wandelt},
  {Wehus}, {White}, {White}, {Wilkinson}, {Yvon}, {Zacchei}, \&
  {Zonca}}]{Ade:2016xua}
{Planck Collaboration}, {Ade}, P.~A.~R., {Aghanim}, N., {et~al.} 2016,
  \href{http://dx.doi.org/10.1051/0004-6361/201525830}{\JournalTitle{\aap},
  594, A13}

\bibitem[{{Regan} {et~al.}(2014){Regan}, {Johansson}, \&
  {Haehnelt}}]{Regan:2014fj}
{Regan}, J.~A., {Johansson}, P.~H., \& {Haehnelt}, M.~G. 2014,
  \href{http://dx.doi.org/10.1093/mnras/stu068}{\JournalTitle{\mnras}, 439,
  1160}

\bibitem[{{Rizzuto} {et~al.}(2020){Rizzuto}, {Naab}, {Spurzem}, {Giersz},
  {Ostriker}, {Stone}, {Wang}, {Berczik}, \& {Rampp}}]{Rizzutto+2020}
{Rizzuto}, F.~P., {Naab}, T., {Spurzem}, R., {et~al.} 2020, \JournalTitle{arXiv
  e-prints}, arXiv:2008.09571

\bibitem[{{Rodriguez} {et~al.}(2020){Rodriguez}, {Kremer}, {Grudi{\'c}},
  {Hafen}, {Chatterjee}, {Fragione}, {Lamberts}, {Martinez}, {Rasio},
  {Weatherford}, \& {Ye}}]{Rodriguez2020}
{Rodriguez}, C.~L., {Kremer}, K., {Grudi{\'c}}, M.~Y., {et~al.} 2020,
  \href{http://dx.doi.org/10.3847/2041-8213/ab961d}{\JournalTitle{\apjl}, 896,
  L10}

\bibitem[{{Roupas} \& {Kazanas}(2019)}]{Roupas2019}
{Roupas}, Z., \& {Kazanas}, D. 2019,
  \href{http://dx.doi.org/10.1051/0004-6361/201937002}{\JournalTitle{\aap},
  632, L8}

\bibitem[{{Ryu} {et~al.}(2016){Ryu}, {Tanaka}, {Perna}, \&
  {Haiman}}]{Ryu:2016ih}
{Ryu}, T., {Tanaka}, T.~L., {Perna}, R., \& {Haiman}, Z. 2016,
  \href{http://dx.doi.org/10.1093/mnras/stw1241}{\JournalTitle{\mnras}, 460,
  4122}

\bibitem[{{Safarzadeh} \& {Berger}(2019)}]{Safarzadeh:2019dn}
{Safarzadeh}, M., \& {Berger}, E. 2019,
  \href{http://dx.doi.org/10.3847/2041-8213/ab24df}{\JournalTitle{\apjl}, 878,
  L12}

\bibitem[{{Safarzadeh} {et~al.}(2019{\natexlab{a}}){Safarzadeh}, {Berger},
  {Leja}, \& {Speagle}}]{Safarzadeh:2019dp}
{Safarzadeh}, M., {Berger}, E., {Leja}, J., \& {Speagle}, J.~S.
  2019{\natexlab{a}},
  \href{http://dx.doi.org/10.3847/2041-8213/ab24e3}{\JournalTitle{\apjl}, 878,
  L14}

\bibitem[{{Safarzadeh} {et~al.}(2019{\natexlab{b}}){Safarzadeh}, {Berger},
  {Ng}, {Chen}, {Vitale}, {Whittle}, \& {Scannapieco}}]{Safarzadeh:2019kj}
{Safarzadeh}, M., {Berger}, E., {Ng}, K. K.~Y., {et~al.} 2019{\natexlab{b}},
  \href{http://dx.doi.org/10.3847/2041-8213/ab22be}{\JournalTitle{\apjl}, 878,
  L13}

\bibitem[{{Sakstein} {et~al.}(2020){Sakstein}, {Croon}, {McDermott},
  {Straight}, \& {Baxter}}]{Sakstein2020}
{Sakstein}, J., {Croon}, D., {McDermott}, S.~D., {Straight}, M.~C., \&
  {Baxter}, E.~J. 2020, \JournalTitle{arXiv e-prints}, arXiv:2009.01213

\bibitem[{{Sakurai} {et~al.}(2016){Sakurai}, {Inayoshi}, \&
  {Haiman}}]{Sakurai+2016}
{Sakurai}, Y., {Inayoshi}, K., \& {Haiman}, Z. 2016,
  \href{http://dx.doi.org/10.1093/mnras/stw1652}{\JournalTitle{\mnras}, 461,
  4496}

\bibitem[{{Salpeter}(1955)}]{Salpeter1955}
{Salpeter}, E.~E. 1955,
  \href{http://dx.doi.org/10.1086/145971}{\JournalTitle{\apj}, 121, 161}

\bibitem[{{Sana} {et~al.}(2012){Sana}, {de Mink}, {de Koter}, {Langer},
  {Evans}, {Gieles}, {Gosset}, {Izzard}, {Le Bouquin}, \&
  {Schneider}}]{Sana2012}
{Sana}, H., {de Mink}, S.~E., {de Koter}, A., {et~al.} 2012,
  \href{http://dx.doi.org/10.1126/science.1223344}{\JournalTitle{Science}, 337,
  444}

\bibitem[{{Shang} {et~al.}(2010){Shang}, {Bryan}, \& {Haiman}}]{Shang+2010}
{Shang}, C., {Bryan}, G.~L., \& {Haiman}, Z. 2010,
  \href{http://dx.doi.org/10.1111/j.1365-2966.2009.15960.x}{\JournalTitle{\mnras},
  402, 1249}

\bibitem[{{Shi} \& {Krolik}(2015)}]{ShiKrolik2015}
{Shi}, J.-M., \& {Krolik}, J.~H. 2015,
  \href{http://dx.doi.org/10.1088/0004-637X/807/2/131}{\JournalTitle{\apj},
  807, 131}

\bibitem[{{Smith} {et~al.}(2018){Smith}, {Regan}, {Downes}, {Norman}, {O'Shea},
  \& {Wise}}]{Smith+2018}
{Smith}, B.~D., {Regan}, J.~A., {Downes}, T.~P., {et~al.} 2018,
  \href{http://dx.doi.org/10.1093/mnras/sty2103}{\JournalTitle{\mnras}, 480,
  3762}

\bibitem[{Tagawa {et~al.}(2019)Tagawa, Haiman, \& Kocsis}]{Tagawa:2019tu}
Tagawa, H., Haiman, Z., \& Kocsis, B. 2019, \JournalTitle{arXiv},
  \href{http://arxiv.org/abs/1912.08218}{{\sffamily 1912.08218}}

\bibitem[{{Tagawa} {et~al.}(2016){Tagawa}, {Umemura}, \& {Gouda}}]{Tagawa2016}
{Tagawa}, H., {Umemura}, M., \& {Gouda}, N. 2016,
  \href{http://dx.doi.org/10.1093/mnras/stw1877}{\JournalTitle{\mnras}, 462,
  3812}

\bibitem[{{Tiede} {et~al.}(2020){Tiede}, {Zrake}, {MacFadyen}, \&
  {Haiman}}]{Tiede2020}
{Tiede}, C., {Zrake}, J., {MacFadyen}, A., \& {Haiman}, Z. 2020,
  \href{http://dx.doi.org/10.3847/1538-4357/aba432}{\JournalTitle{\apj}, 900,
  43}

\bibitem[{{Turk} {et~al.}(2009){Turk}, {Abel}, \& {O'Shea}}]{Turk+2009}
{Turk}, M.~J., {Abel}, T., \& {O'Shea}, B. 2009,
  \href{http://dx.doi.org/10.1126/science.1173540}{\JournalTitle{Science}, 325,
  601}

\bibitem[{Visbal {et~al.}(2020)Visbal, Bryan, \& Haiman}]{Visbal:2020bx}
Visbal, E., Bryan, G.~L., \& Haiman, Z. 2020, \JournalTitle{The Astrophysical
  Journal}

\bibitem[{{Visbal} {et~al.}(2015){Visbal}, {Haiman}, \& {Bryan}}]{Visbal+2015}
{Visbal}, E., {Haiman}, Z., \& {Bryan}, G.~L. 2015,
  \href{http://dx.doi.org/10.1093/mnras/stv1941}{\JournalTitle{\mnras}, 453,
  4456}

\bibitem[{{Whalen} {et~al.}(2004){Whalen}, {Abel}, \& {Norman}}]{Whalen+2004}
{Whalen}, D., {Abel}, T., \& {Norman}, M.~L. 2004,
  \href{http://dx.doi.org/10.1086/421548}{\JournalTitle{\apj}, 610, 14}

\bibitem[{{Whalen} {et~al.}(2008){Whalen}, {van Veelen}, {O'Shea}, \&
  {Norman}}]{Whalen+2008}
{Whalen}, D., {van Veelen}, B., {O'Shea}, B.~W., \& {Norman}, M.~L. 2008,
  \href{http://dx.doi.org/10.1086/589643}{\JournalTitle{\apj}, 682, 49}

\bibitem[{{Wise} {et~al.}(2019){Wise}, {Regan}, {O'Shea}, {Norman}, {Downes},
  \& {Xu}}]{Wise+2019}
{Wise}, J.~H., {Regan}, J.~A., {O'Shea}, B.~W., {et~al.} 2019,
  \href{http://dx.doi.org/10.1038/s41586-019-0873-4}{\JournalTitle{\nat}, 566,
  85}

\bibitem[{{Woosley}(2017)}]{Woosley:2017dj}
{Woosley}, S.~E. 2017,
  \href{http://dx.doi.org/10.3847/1538-4357/836/2/244}{\JournalTitle{\apj},
  836, 244}

\bibitem[{{Yang} {et~al.}(2019){Yang}, {Bartos}, {Gayathri}, {Ford}, {Haiman},
  {Klimenko}, {Kocsis}, {M{\'a}rka}, {M{\'a}rka}, {McKernan}, \&
  {O'Shaughnessy}}]{Yang+2019}
{Yang}, Y., {Bartos}, I., {Gayathri}, V., {et~al.} 2019,
  \href{http://dx.doi.org/10.1103/PhysRevLett.123.181101}{\JournalTitle{\prl},
  123, 181101}

\end{thebibliography}
\end{document}